\documentclass[11pt,a4paper,onecolumn]{article}
\usepackage{graphicx}
\begin{document}
\title{Generalized Statistics and High $T_c$ Superconductivity}

\author{H. Uys \thanks{E-mail:    huys@maple.up.ac.za}, H. G. Miller \thanks{E-Mail:   hmiller@maple.up.ac.za}\\ \textit{\small Department of Physics, University of Pretoria, Pretoria 0002, South Africa \normalsize}}

\date{16 May 2001}

\maketitle

\begin{abstract}
If the generalized statistics suggested by Tsallis are used
in  statistical mechanics,  the fluctuation-dissipation theorem no longer holds . Only in the limiting case where Boltzmann statistics are recovered is the theorem applicable.  In spite of the fact that this allows for the possibility of condensation in any dimension, it is demonstrated that this is not always realized. 
\end{abstract}

In statistical mechanics with the standard Boltzmann entropy, a simple relationship may be obtained between the canonical or grand canonical ensemble averages of commutators and anticommutators of two dynamical operators\cite{KTH95}. This relationship is often referred to as the fluctuation-dissipation theorem since the anticommutator is used to describe time dependent correlations or fluctuations in the system and the commutator is related to transport coefficients or dissipation\cite{N28,CW51}. It is simply due to the fact that the Boltzmann factor or distribution function is exponential in nature and therefore factorizable.
 A consequence of this aforementioned theorem is to  rigorously rule out the existence of superconductivity or superfluidity in one and two dimensions\cite{H67}. 

Recently, however, it has been pointed out by Tsallis\cite{T88} that the Boltzmann-Gibbs statistics may be generalized  such that the entire Legendre-transform structure of  thermodynamics is preserved\cite{CT91}. Although the resulting statistical mechanics is non-extensive, quantal as well as classical applications of the Tsallis statistics have been suggested (see for example the references in Ref~\cite{PPT94}).
 Unlike in the Boltzmann case, however, the generalized distribution function is not simply factorizable, except in the limiting case where Boltzmann-Gibbs statistics are recovered. Hence, as we shall show the fluctuation-dissipation theorem as stated above no longer holds. This allows for the possibility
of forming a condensate in two dimensions, provided these generalized statistics are realized    
as we have suggested is the case of high $T_c$ superconductivity\cite{UMK01}.

In the Tsallis formulation, the entropy is given by
\begin{equation}
S_q=\frac{k_B\Sigma_{m} (\rho_m-\rho^{q-1}_m)}{q-1}
\end{equation}
where $k_B $ is the Boltzmann constant (which will be set equal to 1), q is any real number (characterizing a particular statistics), and the sum runs over all microscopic configurations (whose probabilities are $\{\rho_m\}$) and
\begin{equation}
\Sigma_m \rho_m=1.
\end{equation}
The ensemble average of the internal energy, for example, is given by
\begin{equation}
<E>_q=\Sigma_m \rho_m^q\epsilon_m.
\end{equation}
Furthermore, the generalized Fermi-Dirac (upper sign) and the generalized Bose-Einstein (lower sign) distributions are given by 
\begin{equation}
f_q (\epsilon_k)=\frac{1}{[1+\beta(q-1)(\epsilon_k-\mu)]^{\frac{1}{q-1}}\pm 1} \label{bose}
\end{equation}
and in the Maxwell-Boltzmann case by
\begin{equation}
f_q(\epsilon_k)=[1+\beta(q-1)(\epsilon_k-\mu)]^{\frac{1}{q-1}} \label{MB}
\end{equation}
where $\beta=1/T$ and $\mu $ is the chemical potential \cite{BDG95,PPP95}. In the limit q=1, the standard Boltzmann-Gibbs expressions are recovered.

In the Maxwell-Boltzmann case (with $\mu=0$) consider the simplest form of canonical correlation function\cite{KTH95} 
\begin{equation}
<\hat{A}(t)\hat{B}(t')>=Tr\{\hat{\rho}\hat{A}(t)\hat{B}(t')\}
\end{equation}
where 
\begin{eqnarray}
\hat{\rho}\equiv \hat{\rho}_{q=1}&=&e^{-\beta\hat{H}}/Tr\{e^{-\beta\hat{H}} \}\\
          &=&f_{q=1}(\hat{H})/Z_{q=1}\equiv f(\hat{H})/Z,\label{rho}
\end{eqnarray}
 Z is the partition function and $f_q$ is given by eq.~(\ref{MB}). The spectral function for this correlation function may be written as
\begin{eqnarray}
J_{AB}(\omega)&=&\int^\infty_{-\infty}e^{i\omega t}<\hat{A}(t)\hat{B}(0)> dt\\
      &=&\int\int dE dE'\rho(E)2 \pi \hbar \delta(E+\hbar\omega -E')j_{AB}(E,E')
\end{eqnarray}
where
\begin{equation}
j_{AB}(E,E')=Tr\{\delta(E-\hat{H})\hat{A}\delta(E'-\hat{H})\hat{B}\}
\end{equation}
and
\begin{equation}
\rho(E)=e^{-\beta E}/Z
\end{equation}
which yields for the correlation function
\begin{equation}
<\hat{A}(t)\hat{B}(t')>=\frac{1}{2\pi}\int^\infty_{-\infty}d\omega e^{-i \omega (t-t')} J_{AB}(\omega).
\end{equation}
Interchanging the order of the product and E and E' yields
\begin{equation}
<\hat{B}(t')\hat{A}(t)>=\frac{1}{2\pi}\int^\infty_{-\infty}d\omega e^{-i \omega (t-t')} 
e^{\beta\hbar\omega}J_{AB}(\omega)
\end{equation}
since $\rho(E+\hbar\omega)=\rho(E)e^{-\beta\hbar\omega
}$. This  leads to a simple relationship 
between the ensemble average of the commutator and anticommutator of $\hat{A}$ and $\hat{B}$
which is referred to as the fluctuation-dissipation theorem\cite{KTH95,H67}. Unfortunately this factorization is not possible over the complete integration range
if $\rho$ in eq.~(\ref{rho}) is replaced by $\rho_{q\neq 1}$\cite{PPP95,WM97}.
Hence, in principle, for $q\neq 1$ condensation may occur in dimensions $d\leq 3$.

Consider an ideal Bose gas for which the  number of bosons is given by
\begin{eqnarray}
N(\mu,T)&=& \int\int d^d r d^d p f_q(\epsilon)\\
        &\sim& \int^{\infty}_0 d\epsilon \epsilon^{\frac{d}{2}-1} f_q(\epsilon)
\end{eqnarray}
where d is the dimensionality and $f_q$ is given by eq(~\ref{bose}). For q=1 it is easy to show that $\lim_{\mu \to 0} N_{q=1}(\mu,T)$ is divergent for d=1,2\cite{P72,H67} and condensation only occurs for d=3. On the other hand for q=2,  $\lim_{\mu \to 0} N_{q=2}(\mu,T)$ is not convergent for d=1,2 or 3 which in spite of the absence of the fluctuation-dissipation theorem  rules out the possibility of condensation.

\end{document}